\documentclass[iop]{emulateapj}
\usepackage{color}
\begin{document}

\shorttitle{}
\shortauthors{Falcon, Winget, Montgomery, \& Williams}

\title{A Gravitational Redshift Determination of the Mean Mass of White Dwarfs.  DBA and DB Stars}

\author{Ross~E. Falcon\altaffilmark{1}, D.~E. Winget, M.~H. Montgomery}
\affil{Department of Astronomy and McDonald Observatory, University of Texas, Austin, TX, 78712}
\email{cylver@astro.as.utexas.edu}
\author{and Kurtis~A. Williams}
\affil{Department of Physics and Astronomy, Texas A\&M University-Commerce, Commerce, TX, 75428}
\altaffiltext{1}{National Physical Science Consortium Graduate Fellow}

\begin{abstract}

We measure apparent velocities ($v_{\rm app}$) of absorption lines 
for 36 white dwarfs (WDs) with helium-dominated atmospheres 
-- 16 DBAs and 20 DBs -- using optical spectra taken for the European 
Southern Observatory SN Ia progenitor survey (SPY).   We find a 
difference of $6.9\pm6.9$\,km~s$^{-1}$ in the average apparent 
velocity of the H$\alpha$ lines versus that of the 
\ion{He}{1} 5876\,\AA\ for our DBAs.  This is a measure of the 
blueshift of this He line due to pressure effects.  By using this as 
a correction, we extend the gravitational redshift method employed by 
\citet{Falcon10} to use the apparent velocity of the 
\ion{He}{1} 5876\,\AA\ line and conduct the first gravitational 
redshift investigation of a group of WDs without visible hydrogen lines.  
We use biweight estimators to find an average apparent velocity, 
$\langle v_{\rm app}\rangle_{\rm BI}$, (and hence average gravitational 
redshift, $\langle v_{\rm g}\rangle_{\rm BI}$) for our WDs; from that we 
derive an average mass, $\langle M\rangle_{\rm BI}$.  For the DBAs, we find 
$\langle v_{\rm app}\rangle_{\rm BI}=40.8\pm4.7$\,km~s$^{-1}$ and 
derive $\langle M\rangle_{\rm BI}=0.71^{+0.04}_{-0.05}\,M_\odot$.  
Though different from $\langle v_{\rm app}\rangle$ of DAs 
(32.57 km s$^{-1}$) at the 91\% confidence level and suggestive of a 
larger DBA mean mass than that for normal DAs derived using the same 
method \citep[$0.647^{+0.013}_{-0.014}\,M_\odot$;][]{Falcon10}, we do 
not claim this as a stringent detection.  Rather, we emphasize that the 
difference between $\langle v_{\rm app}\rangle_{\rm BI}$ of the DBAs and 
$\langle v_{\rm app}\rangle$ of normal DAs is no larger than 
9.2\,km s$^{-1}$, at the 95\% confidence level; this corresponds to 
roughly $0.10\,M_\odot$.  For the DBs, we find 
$\langle v_{\rm app}^{\rm He}\rangle_{\rm BI}=42.9\pm8.49$\,km s$^{-1}$ 
after applying the blueshift correction and determine 
$\langle M\rangle_{\rm BI}= 0.74^{+0.08}_{-0.09}\,M_\odot$.  The 
difference between $\langle v_{\rm app}^{\rm He}\rangle_{\rm BI}$ of the 
DBs and $\langle v_{\rm app}\rangle$ of DAs is $\leq11.5$\,km s$^{-1}$ 
($\sim0.12\,M_\odot$), at the 95\% confidence level.  The gravitational 
redshift method indicates much larger mean masses than the spectroscopic 
determinations of the same sample by \citet{Voss07}.  Given the small 
sample sizes, it is possible that systematic uncertainties are skewing 
our results due to the potential of kinematic substructures that may not 
average out.  We estimate this to be unlikely, but a larger sample size 
is necessary to rule out these systematics.

\end{abstract}


\section{Introduction}\label{intro}

In \citet{Falcon10}, we show that the gravitational redshift method is 
an effective tool for measuring mean masses of groups of white dwarfs 
(WDs) and has the advantage of being mostly independent from the 
spectroscopic method \citep[e.g.,][]{Bergeron92b}.  \citet{Falcon10} 
investigate normal DAs, the largest class of WDs.  The next logical step 
is to ask whether the method can be applied to the second largest class, 
DBs, which constitute 20\% of all WDs below $T_{\rm eff}\sim17,000$\,K\, 
and $\sim9$\% of WDs at higher temperatures \citep{Beauchamp96,Bergeron11}.

However, the gravitational redshift method historically uses the apparent 
velocities of hydrogen Balmer line cores.  The work by \citet{Shipman76} 
and by \citet{Grabowski87} show H$\alpha$ to be suitable for this purpose 
since it are not significantly affected by pressure shifts.  Pure DB 
spectra exhibit only helium lines, and as \citet{Greenstein67} first 
pointed out, using WD photospheric helium lines for gravitational redshift 
measurements can be difficult due to the likelihood of systematics 
introduced by pressure effects. These systematics include that, in theory, 
and with some experimental support \citep[e.g.,][]{Berg62,Perez03}, 
different helium lines can be pressure shifted by different amounts, in 
different (blue or red) directions, and with a dependency on temperature 
\citep[e.g.,][]{Griem62,Bassalo76,Dimitrijevic90,Omar06}.  For this 
reason, attempts at gravitational redshift measurements for 
helium-dominated WDs have been sparse.

\citet{Koester87} measures line shifts of \ion{He}{1} 
4026, 4471, 4713, and 4922\,\AA\ in the spectrum of the common proper 
motion star WD~0615-591 and of \ion{He}{1} 4471 and 4922\,\AA\ in the 
wide binary WD~2129+000.  These line shifts are negative (blue), 
meaning that the magnitude of the pressure effects are larger than the 
magnitude of the gravitational redshift, which, in this case, are 
opposing each other.  The fact that these WDs are relatively cool 
\citep{Bergeron11} is consistent with the expectation that 
pressure effects should be significant; we will elaborate on this 
point in Section \ref{heline}.  \citet{Koester87} concludes that 
due to the state of the theory at the time -- laboratory 
measurements and theoretical predictions often disagreed on the 
magnitude and sometimes even sign of the shift -- he cannot deduce 
meaningful gravitational redshifts for these two DBs.


\citet{Wegner89} measure the gravitational redshift for the Hyades 
DBA WD~0437+138 using H$\alpha$ and mention that the velocity ``...is 
unaffected by the pressure shift problems of helium'' while providing 
no further detail.  The importance of van der Waals broadening in cool 
helium-dominated atmospheres \citep{Bergeron91}, however, was perhaps 
not yet well-established.  In hindsight, it is likely that H$\alpha$ 
in this WD {\it is} significantly affected.

The sample of common proper motion binary and cluster WDs in 
\citet{Reid96} contains three targets with helium-dominated atmospheres.  
For both DBAs, WD~0437+138 and WD~1425+540, Reid determines different 
gravitational redshifts from using H$\alpha$ than from H$\beta$.  
He mentions that this discrepancy could be because of pressure shifts 
due to the high atmospheric helium abundance and deems the H$\alpha$ 
result as the better redshift estimate since this line should be less 
affected than H$\beta$.  Reid also measures, like \citet{Koester87}, 
negative shifts of helium lines for the DB WD~2129+000.  One of his 
measured lines (\ion{He}{1} 4921, 5015, and 6678\,\AA) is in common 
with that of \citet{Koester87}.

With more recent high-resolution spectroscopic observations of 
helium-atmosphere WDs \citep{Voss07} and with the method of 
\citet{Falcon10}, we now have the tools to revisit the gravitational 
redshift of DBs.  Such an investigation is a valuable check to the 
latest spectroscopic work \citep{Bergeron11}, the analysis of which 
is nontrivial due in part to the challenge of interpreting 
pressure-broadened helium lines \citep{Beauchamp97,Beauchamp98,Beauchamp99}.

In Section \ref{heline} we discuss using the apparent velocity of the 
\ion{He}{1} 5876\,\AA\ line in the context of our work and, as 
\citet{Wegner87} suggest, check for consistency within the DBA sample 
of the apparent velocities of both the hydrogen and helium line 
species.  In Section \ref{DBAs} we perform the original gravitational 
redshift method that uses hydrogen Balmer lines on the DBAs, which 
results in the most direct comparison with the DAs from \citet{Falcon10}.  
Then in Section \ref{DBs} we extend the method to WDs with only helium 
lines.

\section{Observations}\label{obs}

We use spectroscopic data from the European Southern Observatory (ESO) SN Ia 
progenitor survey \citep[SPY;][]{Napiwotzki01an}.  These observations, taken 
using the UV-Visual Echelle Spectrograph \citep[UVES;][]{Dekker00} at Kueyen, 
Unit Telescope 2 of the ESO VLT array, constitute the largest, homogeneous, 
high-resolution (0.36\,\AA\ or $\sim16$\,km s$^{-1}$ at H$\alpha$) 
spectroscopic dataset for WDs.  We obtain the pipeline-reduced data online 
through the publicly available ESO Science Archive Facility.

\subsection{Samples}\label{sample}


The helium-atmosphere WDs in our sample are derived from the SPY objects
analyzed by \citet{Voss07}.  Out of their 38 DBAs with atmospheric
parameter determinations, we exclude one magnetic WD, one DO, and twelve 
cool WDs with $T_{\rm eff}\lesssim$16,500\,K.  For these twelve, 
\citet{Voss07} fix the surface gravity at log\,$g$=8 in order to obtain
spectral fits; this forfeits our ability to properly compare these stars 
with our gravitational redshift results.  Due to data quality, we are 
unable to satisfactorily measure apparent velocities ($v_{\rm app}$) of 
H$\alpha$ for an additional eight, which leaves us with 16 DBAs.  Keep 
in mind that for these objects, the atmospheric abundance of hydrogen is 
small and H$\alpha$ is often barely visible.

Out of the 31 measured DBs from \citet{Voss07}, we exclude nine cool WDs 
and two known to have cool companions \citep{Zuckerman92,Farihi05,Hoard07}.  
We end up with 20 normal DBs and 36 helium-atmosphere WDs overall.

Four of our DBAs show \ion{Ca}{2} lines in their spectra, giving them 
the additional classification as DBAZ \citep{Koester05.432,Voss07}, and 
two DBs are classified as DBZs \citep{Voss07}.  Since the SPY 
Collaboration has not reported any of our WDs to have stellar companions, 
we presume them to not be in close binary systems, though two of our DBAs 
(WD~0948+013, WD~2154-437) and two of our DBs (WD~0615-591, WD~0845-188) 
are in common proper motion or wide binary systems 
\citep{Luyten49,Wegner73,Caballero09}.  The lack of detectable 
Zeeman splitting implies that our targets also do not harbor significant 
($\gtrsim$100\,kG) magnetic fields \citep[e.g.,][]{Koester09b}.

The WDs in our sample have not been classified in the literature as 
potential members of the thick disk or halo stellar populations.  We 
therefore make the assumption that all of these WDs belong to the thin 
disk -- a necessary assumption for our analysis (see Section \ref{comove}).  
Nearly all ($>$90\%) of the SPY WDs that have been studied kinematically 
are classified as thin disk objects \citep{Pauli06,Richter07}.

\section{Using Gravitational Redshift to Determine a Mean Mass}

We use the methods described in \citet{Falcon10}.  To summarize, we 
measure the apparent velocity of the H$\alpha$ line for an 
ensemble of WDs.  We correct these velocities so that the WDs 
are in a comoving reference frame.  The average apparent velocity 
then becomes the average gravitational redshift because random stellar 
radial velocities average out.  By using the mass-radius relation 
from WD evolutionary models, we translate this average gravitational 
redshift to an average mass.

In this work, we add two extensions to the method: 
(1) a different estimator of central location of the distribution better 
suited for small sample sizes (Section \ref{small}) and
(2) use of the \ion{He}{1} 5876\,\AA\ line for helium-atmosphere WDs 
(Section \ref{heline}).

\subsection{Gravitational Redshift}

In the weak-field limit, the general relativistic effect of gravitational 
redshift can be observed as a velocity shift in absorption lines and is 
expressed as
\begin{equation}\label{v_g}
v_g=\frac{c\Delta\lambda}{\lambda}=\frac{GM}{Rc}
\end{equation}
where $G$ is the gravitational constant, and $c$ is the speed of light.  
In our case, $M$ is the WD mass, and $R$ is the WD radius.

The apparent velocity of an absorption line is the sum of this gravitational 
redshift and the stellar radial velocity: $v_{\rm app}=v_{\rm g}+v_{\rm r}$.  
These two components cannot be explicitly separated for individual WDs 
without an independent $v_{\rm r}$ measurement or mass determination.  We 
break this degeneracy for a group of WDs by assuming that the sample is 
comoving and local.  After we correct each $v_{\rm app}$ to the local 
standard of rest (LSR), only random stellar motions dominate the dynamics 
of our sample.  These average out, and the mean apparent velocity becomes 
the mean gravitational redshift: 
$\langle v_{\rm app}\rangle =\langle v_{\rm g}\rangle$.

\subsection{Velocity Measurements}\label{vel_section}

Collisional (Stark) broadening effects cause asymmetry in the wings of 
absorption lines for the hydrogen Balmer series, making it difficult to 
measure a velocity centroid \citep{Shipman76,Grabowski87}.  However, 
these effects are not significant in the sharp, non-LTE H$\alpha$ and 
H$\beta$ Balmer line cores.  We use both of these line cores in 
\citet{Falcon10}.  In the SPY spectra of DBAs, though, we do not observe 
the non-LTE line cores, but the H$\alpha$ line centers are still distinct.  
For H$\alpha$ the pressure shift should remain very small 
($<1$\,km~s$^{-1}$) within a few \AA\ of the line center \citep{Grabowski87}.  

\citet{Reid96} measures the gravitational redshift of the DBA WD~0437+138 
using the line center since he observes no sharp line core.  While 
cautioning that pressure effects may be significant, he determines a 
high mass of $0.748\pm0.037\,M_\odot$.  \citet{Bergeron11} determine a 
spectroscopic mass of $0.74\pm0.06\,M_\odot$ for this WD.  For at least 
this one case, then, gravitational redshifts of the line center give a 
result consistent with the spectroscopic method.

Van der Waals broadening due to neutral helium may also signficantly 
affect line shapes below $T_{\rm eff}=$16,500\,K 
\citep{Koester05.439,Voss07}.  We avoid this systematic by not including 
targets in that range of $T_{\rm eff}$.

We measure $v_{\rm app}$ for each DBA in our sample by fitting a Gaussian 
profile to the H$\alpha$ line center using GAUSSFIT, a non-linear 
least-squares fitting routine in IDL.  If multiple epochs of observation 
exist, we combine the measurements as a mean weighted according to the 
uncertainties returned by the fitting routine.  Eight out of our 
16 DBAs have two epochs of observation.  Table \ref{table_mix} lists 
these measurements for the DBA sample including those for each epoch of 
observation and the final adopted values.

By inspecting the $v_{\rm app}$ measurements between epochs, we notice they 
differ by more than a typical measurement uncertainty; the mean difference 
between epochs of the same target is 9.20\,km s$^{-1}$ while the mean 
measurement uncertainty for all observations is 3.05\,km s$^{-1}$.  Since we 
presume none of our targets to be in close binary systems 
(Section \ref{sample}), reflex orbital motion cannot be the culprit, and 
therefore it is evident that we are underestimating our individual 
measurement uncertainties.  To more accurately represent our adopted 
$\delta v_{\rm app}$ values for each target, we multiply all observation 
$\delta v_{\rm app}$ values by the factor we find above: the ratio of the 
mean difference in apparent velocity between epochs and the mean apparent 
velocity measurement uncertainty.  For the DBA sample, this factor is 3.02.  
The final adopted values for $\delta v_{\rm app}$ listed in Column 3 in Table 
\ref{table_mix} include this adjustment, and the observation values in Column 
8 are the values before the adjustment.

For the DBAs and the DBs, we also measure $v_{\rm app}$ using the 
\ion{He}{1} 5876\,\AA\ line and list the results in Tables \ref{table_he1} 
and \ref{table_he2}, respectively.  Here, too, we apply a multiplicative 
factor to all observation $\delta v_{\rm app}$ values to better represent 
our measurement uncertainties.  These factors are 2.12 and 2.49 for the two 
samples.  The methodology for line center fitting is the same as used for the 
H$\alpha$ Balmer line.


\subsection{Comoving Approximation}\label{comove}

We measure a mean gravitational redshift by assuming that our WDs are a 
comoving, local sample.  With this assumption, only random stellar motions 
dominate the dynamics of our targets; this falls out when we average over 
the sample.

For this assumption to be valid, at least as an approximation, our WDs must 
belong to the same kinematic population; in our work, this is the thin disk.  
We achieve a comoving group by correcting each measured $v_{\rm app}$ to 
the kinematical LSR described by Standard Solar Motion \citep{Kerr86}.  
Column 4 in Tables \ref{table_mix} - \ref{table_he2} lists the LSR 
corrections applied to each target.  In \citet{Falcon10}, we show in detail 
that this is a suitable choice of reference frame for a sample consisting 
of thin disk WDs.

If the targets in our sample belong predominantly to the thick disk, for 
example, our chosen reference frame will fail to produce a group of objects 
that is at rest with respect to the LSR, introducing a directional bias 
into our measured $\langle v_{\rm app}\rangle$.  This is because the thick 
disk population lags behind the thin disk as it rotates about the Galactic 
center \citep[$\sim40$\,km s$^{-1}$;][]{Gilmore89}.

In contrast to the work on DAs in \citet{Falcon10}, we lack the sample 
size to empirically demonstrate whether or not our WDs move with the LSR.  
Therefore, as we state in Section \ref{sample}, we must assume our targets 
are thin disk objects.

Our WDs must also reside at distances that are small compared to the size 
of the Galaxy so that systematics introduced by the Galactic kinematic 
structure are not significant.  Nine of our DBAs and twelve of our DBs 
have published distance determinations in the range of $\sim25$ to 
$\sim260$\,pc 
\citep{Routly72,Wegner73,Koester81,Farihi05,Castanheira06,Limoges10,Bergeron11}.  
We assume the others are at distances comparable to the rest of the SPY 
sample which \citet{Pauli06} determine spectroscopically to be mostly 
($\gtrsim$90\,\%) within 200\,pc with a mean of $\sim$100\,pc.  Over 
these distances, the velocity dispersion with varying scale height above 
the disk remains modest \citep{Kuijken89}, and differential Galactic 
rotation is negligible \citep{Fich89}.

\subsection{Deriving a Mean Mass}

To translate the mean apparent velocity $\langle v_{\rm app}\rangle$ to 
a mean mass, we invoke two constraints: (1) we need the mass-radius 
relation from an evolutionary model, and (2) since the WD radius does 
slightly contract during its cooling sequence, we need an estimate of the 
position along this track for the average WD in our sample (i.e., a 
mean $T_{\rm eff}$).

Our evolutionary models use $M_{\rm He}/M_\star=10^{-2}$ for the helium 
surface-layer mass, and, since we are interested in WDs with 
helium-dominated atmospheres, we use $M_{\rm H}/M_\star=0$ for the 
hydrogen surface-layer mass.  See \citet{Montgomery99} for a more 
complete description of our models.  Our dependency on evolutionary 
models is small.  We are interested in the mass-radius relation from 
these models, and this is relatively straight-forward since WDs are 
mainly supported by electron degeneracy pressure, making the WD radius 
a weak function of temperature. We estimate that varying the C/O ratio 
in the core affects the radius by less than 0.7\%.

\subsection{Appyling the Method to Small Samples}\label{small}

Besides atmospheric composition, the difference between the WD sample 
in \citet{Falcon10} and the ones in this paper is sample size; the DA 
sample boasts 449 targets while our DBA and DB samples contain 16 and 20, 
respectively.  For large numbers such as for the DAs, the (arithmetic) 
mean is the preferred estimator of central location of the distribution.  
For small numbers, however, this is not necessarily so.

In lieu of the mean, we use the biweight central location estimator and 
the biweight scale estimator as recommended by \citet{Beers90} for a 
kinematic sample of our size.  For a discussion on the biweight and its 
statistical properties of resistance, robustness, and efficiency as they 
relate to sample size and type of distribution, see \citet{Beers90}.

The biweight central location estimator is defined as
\begin{equation}
C_{\rm BI}=M+\frac{\sum_{|u_i|<1}(x_{i}-M)(1-u_i^2)^2}{\sum_{|u_i|<1}(1-u_i^2)^2},
\end{equation}
where $x_i$ are the data, $M$ is the sample median, and $u_i$ are given by
\begin{equation}
u_i=\frac{(x_i-M)}{c\,{\rm MAD}}.
\end{equation}
We set the ``tuning constant'' $c=7.0$ (a slight increase from the 
recommended $c=6.0$ listed in \citet{Beers90}) so that no data rejection 
occurs.  Notice that $C_{\rm BI}$ approaches the arithmetic mean in the 
limit where $c\rightarrow\infty$.  MAD (median absolute deviation from 
the sample median) is defined as
\begin{equation}
{\rm MAD}={\rm median}(|x_i-M|).
\end{equation}
We calculate $C_{\rm BI}$ iteratively, taking $M$ as a first guess, 
substituting the calculated $C_{\rm BI}$ as the improved guess, and 
continuing to convergence.

The biweight scale estimator is defined as
\begin{equation}
S_{\rm BI}=n^{1/2}\frac{\left [\sum_{|u_i|<1}(x_i-M)^2(1-u_i^2)^4\right ]^{1/2}}{\left |\sum_{|u_i|<1}(1-u_i^2)(1-5u_i^2)\right |}.
\end{equation}
Here we set $c=9.0$ (within the expression for $u_i$) as recommended by 
\citet{Beers90}.

To determine the uncertainty of the biweight, we use a Monte Carlo approach.  
We adopt the standard deviation of biweight values from 30,000 simulated 
$v_{\rm app}$ distributions.  Each simulated distribution randomly samples 
from a convolution of two Gaussians.  One represents the measured 
$v_{\rm app}$ distribution by using the biweight scale estimator 
$S_{\rm BI}$ (analogous to standard deviation) of the sample as the width 
of the Gaussian.  The other accounts for the individual apparent velocity 
measurement uncertainties by drawing from a Gaussian with a width equal to 
each $\delta v_{\rm app}$ from the sample.

From here on we use the subscript {\it BI} to denote the biweight central 
location estimator, $C_{\rm BI}$, and biweight scale estimator, 
$S_{\rm BI}$, as used in lieu of the arithmetic mean and standard 
deviation, respectively.

\subsection{Pressure Shifts of Helium Lines}\label{heline}

As described in Section \ref{intro}, helium line centroids are expected to 
shift due to pressure effects, and the magnitude and direction of the shift 
is thought to depend on the specific line.  For that reason, we focus on 
measurements of a single helium line, \ion{He}{1} 5876\,\AA.  Theoretical 
calculations give a negative (blue) Stark shift of the \ion{He}{1} 5876\,\AA\ 
line for electron densities corresponding to WD photospheres and in the 
temperature range relevant to our investigation 
\citep[e.g.,][]{Griem62,Dimitrijevic90}; the experiment by 
\citet{Heading92} confirms the sign of the shift.

\begin{figure}[]
\includegraphics[width=\columnwidth]{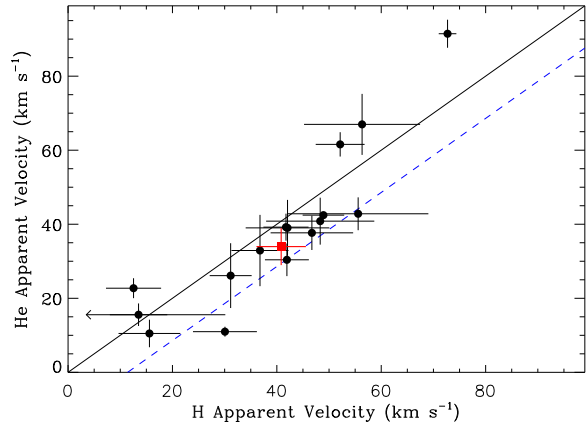}
\caption{Plot of apparent velocity $v_{\rm app}$ obtained from H$\alpha$ 
versus $v_{\rm app}$ obtained from the \ion{He}{1} 5876\,\AA\ line for 
the targets in our DBA sample.  The intersection of 
$\langle v_{\rm app}\rangle_{\rm BI}$ and 
$\langle v_{\rm app}^{\rm He}\rangle_{\rm BI}$ (filled, red square) lies 
below the unity line; the two average values differ by 
$6.9\pm6.9$\,km s$^{-1}$.  The dashed, blue line represents a theoretical 
blueshift of the \ion{He}{1} line derived from \citet{Dimitrijevic90} 
for $T=16,500$\,K and $n_{\rm e}=10^{17}$\,cm$^{-3}$ (see Section 
\ref{heline}).
\label{vel-vel}}
\end{figure}

We search for this expected blueshift using our sample of DBAs, whose
spectra show both the \ion{He}{1} 5876\,\AA\ line and the H$\alpha$ Balmer 
line center that should not be affected by pressure shifts.  
Figure \ref{vel-vel} shows a comparison of apparent velocities measured 
from these lines, and indeed the intersection (filled, red square) of 
the average (biweight) values lies below the unity line.

Using the calculated shifts from \citet{Dimitrijevic90}, we plot a 
dashed line in Figure \ref{vel-vel} indicating the deviation from unity 
for plasma conditions corresponding to temperature $T=16,500$\,K and 
electron density $n_{\rm e}=10^{17}$\,cm$^{-3}$.  This should represent 
an overestimate of the expected shift based on theory.  
$T_{\rm eff}=16,500$\,K is the lower limit for the WDs in our sample, and 
not only does the magnitude of the shift increase with decreasing 
temperature, but it tracks exponentially so that the effect is more 
exaggerated at lower $T$.  Furthermore, although
$n_{\rm e}=10^{17}$\,cm$^{-3}$ is typical for WD photospheres, we are 
concerned with absorption line centers.  These are formed higher 
in the atmosphere where the density is lower and the shift should be less.  
As expected, the intersection of our measured biweight values lies between 
the unity line and the blueshift overestimate.

The presence of hydrogen in a helium-dominated atmosphere may also affect 
the atmospheric pressure and therefore electron density because of its 
relatively high opacity; in our temperature range, the opacity of hydrogen 
can be an order of magnitude greater than that of helium.  From 
\citet{Voss07}, the mean atmospheric hydrogen abundance for the WDs in our 
DBA sample is H/He= $10^{-3.29}$ with the highest abundance of any WD 
being H/He= $10^{-2}$.  At these abundances, this opacity effect should be 
negligible, and the dependency of the distribution of pressure shifts on 
electron density should only be due to the intrinsic mass distribution.

The average apparent velocity of the DBAs in Figure \ref{vel-vel} as 
measured from H$\alpha$ is 
$\langle v_{\rm app}\rangle_{\rm BI}=40.8\pm4.7$\,km s$^{-1}$; 
the average velocity as measured from the \ion{He}{1} 5876\,\AA\ line is 
$\langle v_{\rm app}^{\rm He}\rangle_{\rm BI}=34.0\pm5.0$\,km s$^{-1}$.  
The measured difference of $6.9\pm6.9$\,km s$^{-1}$ is blueshifted and 
exactly at the 1-sigma boundary of our measurement precision.

With assumptions this measured blueshift can be used as a correction 
to apply to the average apparent velocity of DB WD samples in order 
to estimate an average gravitational redshift.  First, one must assume 
that DBAs are not fundamentally different from DBs in a way that would 
manifest as different mean masses.  \citet{Voss07} detect various 
amounts of hydrogen in most (55\%) of the helium-dominated WDs in their 
sample and find similar spectroscopic mass distributions between the 
DBAs and DBs.  \citet{Bergeron11} also find no significant differences 
between the masses of the two groups.  This provides evidence to the 
idea of \citet{Weidemann91} that the DBA subclass is not distinct from 
its parent class but rather the observationally detectable end of a 
continuous distribution of hydrogen abundances.

Also, though our value is indeed a measure of the average blueshift 
due to pressure effects, this average is not straight-forward.  The 
magnitude and sign of the shift depend on temperature and electron 
density (or pressure) \citep[e.g.,][]{Kobilarov89,Dimitrijevic90}, and 
the atmospheres of our DBAs sample a distribution of these plasma 
conditions.  For our measured value to be representative of the average 
blueshift for another sample of WDs, it is important that this sample 
have a similar $T_{\rm eff}$ distribution.  Assessing the similarity 
between samples when applying this correction will deserve further 
scrutiny when a future analysis is performed which uses significantly 
larger sample sizes and hence higher precision.

We apply this correction so that we may, for the first time, estimate 
an average gravitational redshift for DB WDs in the field.  The 
precision of this correction is given by the combined uncertainty of 
the method using the H$\alpha$ Balmer line and of the one using the He 
line (i.e., $\delta\langle v_{\rm app}\rangle_{\rm BI}$ and 
$\delta\langle v_{\rm app}^{\rm He}\rangle_{\rm BI}$ added in 
quadrature; $\sim7$\,km s$^{-1}$) and can be improved by increasing 
the sample size of DBA WDs.

\section{Results}\label{results}

\subsection{DBAs}\label{DBAs}

\subsubsection{Analysis with the H$\alpha$ Balmer Line}

\begin{figure}[]
\includegraphics[width=\columnwidth]{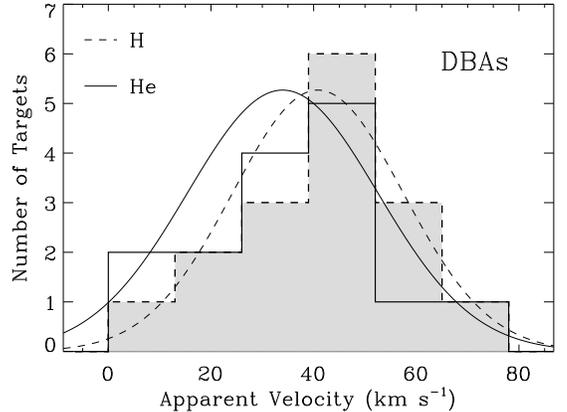}
\caption{Histograms of measured apparent velocities of H$\alpha$ 
$v_{\rm app}$ (dashed, filled) and of \ion{He}{1} 5876\,\AA\ 
$v_{\rm app}^{\rm He}$ (solid, unfilled) for 16 DBAs.  The bin size is 
13\,km~s$^{-1}$.  The dashed and solid curves are the Gaussian 
distributions used to determine Monte Carlo uncertainties for each 
sample.  For H, 
$\langle v_{\rm app}\rangle_{\rm BI}=40.8\pm4.7$\,km s$^{-1}$ 
and $\sigma(v_{\rm app})_{\rm BI}=16.6$\,km s$^{-1}$.  For He, 
$\langle v_{\rm app}^{\rm He}\rangle_{\rm BI}=34.0\pm5.0$\,km s$^{-1}$ 
and $\sigma(v_{\rm app}^{\rm He})_{\rm BI}=18.6$\,km s$^{-1}$.
\label{vdist_mix}}
\end{figure}

We observe hydrogen absorption in 16 WDs classifed as helium-dominated by 
\citet{Voss07}.  Figure \ref{vdist_mix} shows the distribution of 
$v_{\rm app}$ measured from the H$\alpha$ Balmer line.  Table 
\ref{table_mix} lists these measured values, which are described in 
Section \ref{vel_section}.

For this sample, 
$\langle v_{\rm app}\rangle_{\rm BI}=40.8\pm4.7$\,km s$^{-1}$.  We find 
that this is different from $\langle v_{\rm app}\rangle$ of DAs 
(32.57\,km s$^{-1}$) at the 91\% confidence level.

Given the small sample size, it is possible that a systematic uncertainty 
is skewing our result.  For example, by chance kinematic substructures 
may not average out with only 16 targets as they would for a larger 
sample.  As a check, we perform the following test: let us assume that 
the SPY DBAs have the same intrinsic mean mass as the SPY DAs.  If we 
randomly pick 16 apparent velocities from the measured $v_{\rm app}$ 
distribution from \citet{Falcon10}, which contains 449 objects, how often 
is the $\langle v_{\rm app}\rangle_{\rm BI}$ of these 16 greater than 
$\langle v_{\rm app}\rangle_{\rm BI}=40.8$\,km s$^{-1}$, our result 
for DBAs?  Performing the random selection 30,000 times, we find that 
this false detection occurs 9\% of the time.  The synthetic biweights 
reproduce the $\langle v_{\rm app}\rangle$ of DAs with a standard 
deviation of 4.17\,km~s$^{-1}$.  There is a rare but not insignificant 
possibility that our small sample size is fooling us.

Using spectroscopically determined $T_{\rm eff}$ from \citet{Voss07}, 
$\langle T_{\rm eff}\rangle_{\rm BI}=17,890\pm250$\,K.  We use the 
mass-radius relation from our evolutionary models and interpolate to find 
$\langle M\rangle_{\rm BI}=0.71^{+0.04}_{-0.05}\,M_\odot$.  This is 
larger than the mean mass for normal DAs determined using the same 
gravitational redshift method 
\citep[$0.647^{+0.013}_{-0.014}\,M_\odot$;][]{Falcon10}.

Though suggestive of a real difference in mean mass between the samples, 
this is not a stringent result.  It remains plausible that no 
intrinsic difference exists.  We can say, however, that the difference 
between $\langle v_{\rm app}\rangle_{\rm BI}$ of the DBAs and 
$\langle v_{\rm app}\rangle$ of normal DAs is no larger than 
9.2\,km s$^{-1}$, at the 95\% confidence level.  With a note of caution, 
because of the subleties in translating a mean apparent velocity to a 
mean mass, this corresponds to roughly $0.10\,M_\odot$.

Our $\langle M\rangle_{\rm BI}$ for DBAs is also larger than 
$0.62\pm0.02\,M_\odot$, the biweight of the spectroscopic mass 
determinations for these targets from \citet{Voss07}.  The value for the 
mean is the same as the biweight.  We estimate the uncertainty using the 
same method as for our work, except that, since \citet{Voss07} do not 
list individual uncertainties, we assign each mass an uncertainty equal 
to the standard deviation of the mass distribution.  The biweight (and 
mean) spectroscopic mass of the 22 DBAs with $T_{\rm eff}\ge16,500$\,K 
from \citet{Bergeron11} agrees well at $0.67\pm0.02\,M_\odot$.  For this 
value we use the mass uncertainties from \citet{Bergeron11}.


\subsubsection{Analysis with the \ion{He}{1} 5876\,\AA\ Line}\label{withhe}

By measuring $v_{\rm app}^{\rm He}$ -- instead of that from the 
H$\alpha$ line -- for the targets in our DBA sample, we find 
$\langle v_{\rm app}^{\rm He}\rangle_{\rm BI}=34.0\pm5.0$\,km s$^{-1}$.  
Table \ref{table_he1} lists individual measurements.  As discussed in 
Section \ref{heline}, this differs from the H$\alpha$ result by 
$6.9\pm6.9$\,km~s$^{-1}$.  It lies between the unity line and the 
theoretical blueshift derived from \citet{Dimitrijevic90} for plasma 
conditions corresponding to $T=16,500$\,K and 
$n_{\rm e}=10^{17}$\,cm$^{-3}$.  This temperature is the lower limit of 
our sample, and this electron density is typical of WD photospheres.

\subsection{DBs}\label{DBs}

We now use the \ion{He}{1} 5876\,\AA\ line to perform the gravitational 
redshift method on DBs.  We measure $v_{\rm app}^{\rm He}$ for 
20 such WDs and find 
$\langle v_{\rm app}^{\rm He}\rangle_{\rm BI}=36.0\pm5.6$\,km s$^{-1}$.  
Table \ref{table_he2} lists individual measurements.  After applying 
the correction for the blueshift due to pressure effects measured in 
Section \ref{heline}, 
$\langle v_{\rm app}^{\rm He}\rangle_{\rm BI}=42.9\pm8.9$\,km s$^{-1}$.  
This uncertainty comes from adding the measured uncertainty and that of 
the correction in quadrature.  Using 
$\langle T_{\rm eff}\rangle_{\rm BI}=20,570\pm660$\,K \citep{Voss07} 
with the corrected $\langle v_{\rm app}^{\rm He}\rangle_{\rm BI}$, we 
determine $\langle M\rangle_{\rm BI}=0.74^{+0.08}_{-0.09}\,M_\odot$.  

This is different from $\langle v_{\rm app}\rangle$ of DAs at the 75\% 
confidence level. The difference between 
$\langle v_{\rm app}^{\rm He}\rangle_{\rm BI}$ (with the blueshift 
correction) of the DBs and $\langle v_{\rm app}\rangle$ of DAs is 
$\leq11.5$\,km s$^{-1}$, at the 95\% confidence level.  This 
corresponds to roughly $0.12\,M_\odot$.

From \citet{Voss07}, we find 
$\langle M^*\rangle_{\rm BI}=0.58\pm0.02\,M_\odot$ for 
these targets ($M^*$ denotes spectroscopic mass); the mean is 
$0.59\pm0.02\,M_\odot$.  The biweight 
spectroscopic mass of the 35 DBs with $T_{\rm eff}\ge16,500$\,K from 
\citet{Bergeron11} is $0.63\pm0.01\,M_\odot$; the mean is 
$0.64\pm0.01\,M_\odot$. The mean 
of the 82 DBs with $T_{\rm eff}>16,000$\,K from SDSS and whose spectra 
have S/N$\geq20$ is $0.646\pm0.006\,M_\odot$ \citep{Kepler10}.

\subsection{DBAs and DBs}

As mentioned in Section \ref{heline}, let us assume there are no fundamental 
differences between DBAs and DBs that would manifest as different mean 
masses.  We can then apply the correction for the blueshift due to 
pressure effects, and we can improve both the precision and accuracy of our 
mean value determinations by combining the two samples.  We do not discard 
the possibility, however, that the two groups are indeed fundamentally 
different.

\begin{figure}
\includegraphics[width=\columnwidth]{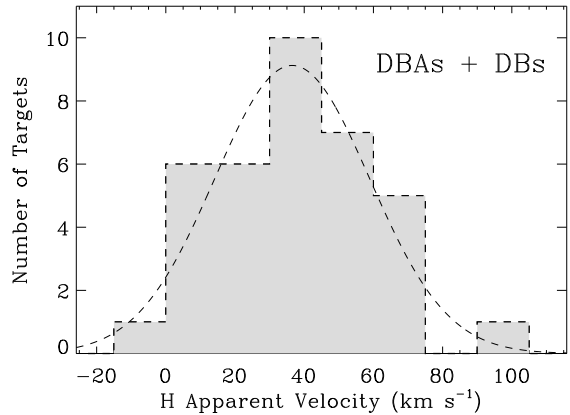}
\caption{Histogram of measured apparent velocities of the \ion{He}{1} 
5876\,\AA\ line $v_{\rm app}^{\rm He}$ for 36 DBAs and DBs (16 and 20, 
respectively) with a bin size is 15\,km~s$^{-1}$.  The dashed curve 
is the Gaussian distribution used to determine Monte Carlo 
uncertainties.  
$\langle v_{\rm app}^{\rm He}\rangle_{\rm BI}=36.5\pm3.9$\,km s$^{-1}$ 
(before applying the correction discussed in Section \ref{heline}) and 
$\sigma (v_{\rm app}^{\rm He})_{\rm BI}=22.4$\,km s$^{-1}$.
\label{vdist_db_plus}}
\end{figure}

Using the apparent velocity of the \ion{He}{1} 5876\,\AA\ line for the 
combined sample of DBAs and DBs, we find 
$\langle v_{\rm app}^{\rm He}\rangle_{\rm BI}=36.5\pm3.9$\,km s$^{-1}$.  
Figure \ref{vdist_db_plus} shows the distribution of $v_{\rm app}^{\rm He}$.  
Applying the blueshift correction, 
$\langle v_{\rm app}^{\rm He}\rangle_{\rm BI}=43.4\pm7.9$\,km s$^{-1}$.
Using $\langle T_{\rm eff}\rangle_{\rm BI}=19,260\pm390$\,K, we determine 
$\langle M\rangle_{\rm BI}=0.74^{+0.07}_{-0.08}\,M_\odot$.

This is different from the mean $v_{\rm app}$ of DAs at the 82\% 
confidence level.  The difference between 
$\langle v_{\rm app}^{\rm He}\rangle_{\rm BI}$ of this combined sample of 
DBAs + DBs and $\langle v_{\rm app}\rangle$ of DAs is 
$\leq7.7$\,km s$^{-1}$, at the 95\% confidence level.  This corresponds to 
roughly $0.08\,M_\odot$.

$\langle M^*\rangle_{\rm BI}=0.60\pm0.02\,M_\odot$ for these 
targets using the spectroscopic masses from \citet{Voss07}.  This is 
significantly lower than our determination.  The biweight (and mean) 
spectroscopic mass of the 57 DBAs and DBs with $T_{\rm eff}\ge16,500$\,K 
from \citet{Bergeron11} is $0.65\pm0.01\,M_\odot$.

We list our derived $\langle v_{\rm app}\rangle_{\rm BI}$ in Table 
\ref{results_mix}, denoting which samples use the \ion{He}{1} 5876\,\AA\ 
line in lieu of the H$\alpha$ Balmer line.  Since the 
$\langle v_{\rm app}\rangle_{\rm BI}$ values are entirely model independent, 
we list them apart from the $\langle M\rangle_{\rm BI}$ we determine.  These 
are in Table \ref{table_mass}.  For comparison, we list the corresponding 
results for DAs from \citet{Falcon10}.

\section{Conclusions}

We measure the apparent velocity ($v_{\rm app}$) of the 
\ion{He}{1} 5876\,\AA\ line for a sample of DBAs and compare it to 
that of the H$\alpha$ Balmer line.  We find a difference of 
$6.9\pm6.9$\,km~s$^{-1}$ in the average apparent velocities from the two 
line species, which we attribute to the blueshift of this He line due 
to pressure effects \citep[e.g.,][]{Dimitrijevic90} averaged over the 
sample.  With assumptions one can apply this measured blueshift as a 
correction to other samples of helium-atmosphere WDs.  We do so in order 
to investigate the average gravitational redshift of a sample of DBs 
for the first time.

Following the gravitational redshift method from \citet{Falcon10}, 
but using biweight estimators which are better suited for small 
sample sizes, we find 
$\langle v_{\rm g}\rangle_{\rm BI}= \langle v_{\rm app}\rangle_{\rm BI}= 40.8\pm4.7$\,km~s$^{-1}$ 
for 16 DBAs with $T_{\rm eff}\ge16,500$\,K from SPY.  We translate this 
$\langle v_{\rm app}\rangle_{\rm BI}$ to a mass: 
$\langle M\rangle_{\rm BI} = 0.71^{+0.04}_{-0.05}\,M_\odot$.  Though 
different from the $\langle v_{\rm app}\rangle$ of DAs, 32.57 km s$^{-1}$, 
at the 91\% confidence level and suggestive of a larger DBA mean mass than 
DA, $0.647^{+0.013}_{-0.014}\,M_\odot$ \citep{Falcon10}, this is not a 
stringent result.  We emphasize that the difference between 
$\langle v_{\rm app}\rangle_{\rm BI}$ of the DBAs and 
$\langle v_{\rm app}\rangle$ of normal DAs is no larger than 
9.2\,km~s$^{-1}$, at the 95\% confidence level; this corresponds to 
roughly $0.10\,M_\odot$.  Our $\langle M\rangle_{\rm BI}$ for DBAs is 
also larger than the average of the spectroscopic mass determinations for 
these targets from \citet{Voss07} at $0.62\pm0.02\,M_\odot$.  It agrees 
with the average mass of the 22 DBAs with $T_{\rm eff}\ge16,500$\,K from 
\citet{Bergeron11} at $0.67\pm0.02\,M_\odot$.

We use the \ion{He}{1} 5876\,\AA\ line to conduct the first gravitational 
redshift investigation of a group of WDs without visible hydrogen lines.  
For 20 DBs from SPY, we find 
$\langle v_{\rm app}^{\rm He}\rangle_{\rm BI}=42.9\pm8.9$\,km s$^{-1}$ 
after applying the correction for our measured blueshift due to pressure 
effects.  We determine 
$\langle M\rangle_{\rm BI}= 0.74^{+0.08}_{-0.09}\,M_\odot$.  The 
difference between $\langle v_{\rm app}^{\rm He}\rangle_{\rm BI}$ of the 
DBs and $\langle v_{\rm app}\rangle$ of DAs is $\leq11.5$\,km s$^{-1}$, 
at the 95\% confidence level; this corresponds to roughly $0.12\,M_\odot$.  
The $\langle M\rangle_{\rm BI}$ is much larger than the average of the 
spectroscopic mass determinations from SPY, $0.58\pm0.02\,M_\odot$.  It 
is slightly larger than the average spectroscopic mass of DBs with 
$T_{\rm eff}>16,500$\,K from \citet{Bergeron11}, 
$0.63\pm0.01\,M_\odot$, and the mean of DBs with $T_{\rm eff}>16,000$\,K 
from SDSS, $0.646\pm0.006\,M_\odot$ \citep{Kepler10}.

Combining our DBA and DB samples to group all WDs with helium-dominated 
atmospheres, we find 
$\langle v_{\rm app}^{\rm He}\rangle_{\rm BI}= 43.4\pm7.9$\,km s$^{-1}$ 
(after the correction) and determine 
$\langle M\rangle_{\rm BI}= 0.74^{+0.07}_{-0.08}\,M_\odot$.  This differs 
from the $\langle v_{\rm app}\rangle$ of DAs at the 82\% confidence 
level.  The difference between 
$\langle v_{\rm app}^{\rm He}\rangle_{\rm BI}$ of the DBAs + DBs and 
$\langle v_{\rm app}\rangle$ of DAs is $\leq7.7$\,km~s$^{-1}$ (roughly 
$0.08\,M_\odot$), at the 95\% confidence level.  Our 
$\langle M\rangle_{\rm BI}$ is much larger than the average 
spectroscopic mass of these targets from SPY at $0.60\pm0.02\,M_\odot$ 
and slightly larger than the average spectroscopic mass of the DBAs and 
DBs with $T_{\rm eff}\ge16,500$\,K from \citet{Bergeron11}, 
$0.65\pm0.01\,M_\odot$.

Given the small sample sizes, it is possible that system uncertainties 
are skewing our results due to the potential of kinematic substructures 
that may not average out.  We estimate this to be unlikely, but a 
larger sample size is necessary to rule out these systematics.

\acknowledgements 
We thank Pierre Bergeron for valuable comments.  R.E.F. thanks the SPY 
Collaboration for providing reduced spectra of excellent quality.  The 
observations were made with the European Southern Observatory telescopes 
and obtained from the ESO/ST-ECF Science Archive Facility.  This work has 
made use of NASA's Astrophysics Data System Bibliographic Services. It 
has also made use of the SIMBAD database, operated at CDS, Strasbourg, 
France.  This work is supported by the National Science Foundation under 
grant AST-0909107, the Norman Hackerman Advanced Research Program under 
grants 003658-255-2007 and 003658-0252-2009, and the Institute for High 
Energy Density Science, funded by The University of Texas System and 
supported in part by Sandia National Laboratories.  R.E.F. acknowledges 
support of the National Physical Science Consortium.  M.H.M acknowledges 
the support of the Delaware Asteroseismic Research Center.  K.A.W. 
acknowledges the additional financial support of NSF award AST-0602288 
and NASA grants NNX11AG82G and HST-GO-11141.

\bibliographystyle{apj}
\bibliography{/home/grad79/cylver/all}

\begin{deluxetable}{cccccccc}
\tablewidth{0pt}
\tabletypesize{\footnotesize}
\tablecaption{H Apparent Velocity Measurements for DBAs\label{table_mix}}
\tablehead{
\colhead{Target} & \multicolumn{2}{c}{Adopted} & \colhead{Date} & \colhead{Time} & \colhead{LSR} & 
\multicolumn{2}{c}{Observation}
\\
\cline{2-3} \cline{7-8} \vspace{-0.12in}
\\
\colhead{} & \colhead{$v_{\rm app}$} & \colhead{$\delta v_{\rm app}$} & \colhead{} & \colhead{} & \colhead{Correction} &  \colhead{$v_{\rm app}$} & \colhead{$\delta v_{\rm app}$} 
\\
\colhead{} & \colhead{(km s$^{-1}$)} & \colhead{(km s$^{-1}$)} & \colhead{(UT)} & \colhead{(UT)} & \colhead{(km s$^{-1}$)} &  \colhead{(km s$^{-1}$)} & \colhead{(km s$^{-1}$)}
}
\startdata
HE 0025-0317  & 36.76 &  5.45  & 2000.07.17 & 07:52:34 &  27.319 & 36.76 & 1.80  \\
HE 0110-5630  & 42.04 &  8.01  & 2002.09.25 & 07:28:16 & -16.836 & 42.04 & 2.65  \\
WD 0125-236   & 55.57 & 13.45  & 2002.09.30 & 04:23:09 &  -6.499 & 55.57 & 4.45  \\ 
WD 0921+091   & 46.71 &  7.90  & 2001.04.07 & 00:55:54 & -35.250 & 53.62 & 2.49  \\
              &       &        & 2002.12.28 & 07:48:25 &  10.131 & 42.19 & 2.01  \\
WD 0948+013   & 30.05 &  6.12  & 2001.01.10 & 01:31:21 & -34.417 & 38.21 & 4.20  \\
              &       &        & 2003.01.17 & 04:03:53 &   4.365 & 27.75 & 2.23  \\
WD 1149-133   & 48.92 &  3.98  & 2000.07.13 & 23:40:32 & -34.214 & 53.48 & 3.71  \\
              &       &        & 2000.07.16 & 23:55:06 & -33.761 & 47.18 & 2.29  \\
HE 1207-2349  & 31.14 &  4.02  & 2002.02.23 & 07:41:20 &  12.910 & 43.82 & 6.79  \\
              &       &        & 2002.02.24 & 07:11:53 &  12.571 & 30.51 & 1.52  \\
EC 12438-1346 & 13.50 & 16.32  & 2000.07.15 & 00:27:35 & -31.791 & 13.50 & 5.80  \\
WD 1311+129   & 12.54 &  5.28  & 2001.06.20 & 01:47:22 & -20.994 & 14.87 & 2.87  \\
              &       &        & 2003.01.18 & 08:36:37 &  33.896 &  6.55 & 4.60  \\
HE 1349-2305  & 15.58 &  5.91  & 2000.07.15 & 02:59:13 & -28.698 & 15.58 & 1.95  \\
WD 1421-011   & 56.32 & 11.11  & 2001.08.16 & 00:32:05 & -19.271 & 56.32 & 3.67  \\
WD 1557+192   & 52.12 &  4.68  & 2002.04.23 & 06:37:23 &  22.663 & 52.12 & 1.55  \\
WD 1709+230   & 72.70 &  1.68  & 2002.09.04 & 00:10:33 &  -1.552 & 71.41 & 3.30  \\
              &       &        & 2002.09.21 & 00:32:10 &  -1.091 & 73.80 & 3.05  \\
WD 2154-437   & 41.92 &  4.20  & 2000.06.04 & 06:01:21 &  25.025 & 41.92 & 1.39  \\
WD 2253-062   & 41.47 &  4.32  & 2000.06.01 & 08:50:29 &  37.290 & 37.64 & 3.38  \\
              &       &        & 2000.06.06 & 08:07:23 &  37.293 & 44.02 & 2.51  \\
HE 2334-4127  & 48.30 & 10.36  & 2002.09.14 & 02:03:49 &  -9.502 & 54.14 & 2.13  \\
              &       &        & 2002.09.15 & 01:27:02 &  -9.859 & 39.11 & 2.67 
\enddata
\end{deluxetable}

\begin{deluxetable}{cccccccc}
\tablewidth{0pt}
\tabletypesize{\footnotesize}
\tablecaption{He Apparent Velocity Measurements for DBAs\label{table_he1}}
\tablehead{
\colhead{Target} & \multicolumn{2}{c}{Adopted} & \colhead{Date} & \colhead{Time} & \colhead{LSR} & \multicolumn{2}{c}{Observation}
\\
\cline{2-3} \cline{7-8} \vspace{-0.12in}
\\
\colhead{} & \colhead{$v_{\rm app}^{\rm He}$} & \colhead{$\delta v_{\rm app}^{\rm He}$} & \colhead{} & \colhead{} & \colhead{Correction} & \colhead{$v_{\rm app}^{\rm He}$} & \colhead{$\delta v_{\rm app}^{\rm He}$} 
\\
\colhead{} & \colhead{(km s$^{-1}$)} & \colhead{(km s$^{-1}$)} & \colhead{(UT)} & \colhead{(UT)} & \colhead{(km s$^{-1}$)} & \colhead{(km s$^{-1}$)} & \colhead{(km s$^{-1}$)}
}
\startdata
HE 0025-0317  & 32.90 & 9.63  & 2000.07.17 & 07:52:34 &  27.319 & 32.90 & 4.54   \\
HE 0110-5630  & 39.01 & 7.55  & 2002.09.25 & 07:28:16 & -16.836 & 39.01 & 3.56   \\
WD 0125-236   & 42.83 & 4.43  & 2002.09.30 & 04:23:09 &  -6.499 & 42.83 & 2.09   \\ 
WD 0921+091   & 37.67 & 4.63  & 2001.04.07 & 00:55:54 & -35.250 & 40.38 & 2.55   \\
              &       &       & 2002.12.28 & 07:48:25 &  10.131 & 33.71 & 3.08   \\
WD 0948+013   & 10.98 & 1.33  & 2001.01.10 & 01:31:21 & -34.417 & 12.76 & 2.76   \\
              &       &       & 2003.01.17 & 04:03:53 &   4.365 & 10.48 & 1.46   \\
WD 1149-133   & 42.45 & 0.18  & 2000.07.13 & 23:40:32 & -34.214 & 42.13 & 6.68   \\
              &       &       & 2000.07.16 & 23:55:06 & -33.761 & 42.51 & 2.68   \\
HE 1207-2349  & 26.12 & 8.75  & 2002.02.23 & 07:41:20 &  12.910 & 20.72 & 3.80   \\
              &       &       & 2002.02.24 & 07:11:53 &  12.571 & 33.22 & 4.36   \\
EC 12438-1346 & 15.56 & 3.04  & 2000.07.15 & 00:27:35 & -31.791 & 15.56 & 1.43   \\
WD 1311+129   & 22.72 & 2.69  & 2001.06.20 & 01:47:22 & -20.994 & 21.59 & 2.15   \\
              &       &       & 2003.01.18 & 08:36:37 &  33.896 & 25.91 & 3.61   \\
HE 1349-2305  & 10.49 & 3.72  & 2000.07.15 & 02:59:13 & -28.698 & 10.49 & 1.75   \\
WD 1421-011   & 66.97 & 8.23  & 2001.08.16 & 00:32:05 & -19.271 & 66.97 & 3.88   \\
WD 1557+192   & 61.57 & 3.28  & 2002.04.23 & 06:37:23 &  22.663 & 61.57 & 1.55   \\
WD 1709+230   & 91.46 & 3.78  & 2002.09.04 & 00:10:33 &  -1.552 & 89.91 & 1.58   \\
              &       &       & 2002.09.21 & 00:32:10 &  -1.091 & 96.09 & 2.73   \\
WD 2154-437   & 30.39 & 4.40  & 2000.06.04 & 06:01:21 &  25.025 & 30.39 & 2.07   \\
WD 2253-062   & 39.21 & 3.64  & 2000.06.01 & 08:50:29 &  37.290 & 37.50 & 1.36   \\
              &       &       & 2000.06.06 & 08:07:23 &  37.293 & 43.11 & 2.06   \\
HE 2334-4127  & 40.83 & 6.33  & 2002.09.14 & 02:03:49 &  -9.502 & 37.32 & 2.17   \\
              &       &       & 2002.09.15 & 01:27:02 &  -9.859 & 46.56 & 2.77
\enddata
\end{deluxetable}

\begin{deluxetable}{cccccccc}
\tablewidth{0pt}
\tabletypesize{\footnotesize}
\tablecaption{He Apparent Velocity Measurements for DBs\label{table_he2}}
\tablehead{
\colhead{Target} & \multicolumn{2}{c}{Adopted} & \colhead{Date} & \colhead{Time} & \colhead{LSR} & \multicolumn{2}{c}{Observation}
\\
\cline{2-3} \cline{7-8} \vspace{-0.12in}
\\
\colhead{} & \colhead{$v_{\rm app}^{\rm He}$} & \colhead{$\delta v_{\rm app}^{\rm He}$} & \colhead{} & \colhead{} & \colhead{Correction} & \colhead{$v_{\rm app}^{\rm He}$} & \colhead{$\delta v_{\rm app}^{\rm He}$} 
\\
\colhead{} & \colhead{(km s$^{-1}$)} & \colhead{(km s$^{-1}$)} & \colhead{(UT)} & \colhead{(UT)} & \colhead{(km s$^{-1}$)} & \colhead{(km s$^{-1}$)} & \colhead{(km s$^{-1}$)}
}
\startdata
MCT 0149-2518 & 62.43 &  7.83  & 2002.09.27 & 08:32:10 &  -3.801 & 55.45 & 2.56   \\
              &       &        & 2002.09.30 & 04:09:09 &  -4.575 & 66.83 & 2.03   \\
WD 0158-160   & 47.19 &  3.20  & 2002.09.19 & 03:40:58 &   0.897 & 48.67 & 1.61   \\
              &       &        & 2002.09.20 & 03:55:16 &   0.462 & 43.71 & 2.46   \\
WD 0249-052   & 46.41 &  2.16  & 2002.09.20 & 07:09:51 &   9.064 & 50.32 & 3.97   \\
              &       &        & 2002.12.31 & 02:03:00 & -33.912 & 45.81 & 1.55   \\
HE 0308-5635  & 63.79 &  6.87  & 2002.09.14 & 05:54:31 & -12.226 & 63.79 & 2.75   \\
WD 0349+015   &  8.96 &  2.21  & 2003.01.24 & 03:40:30 & -38.561 &  8.96 & 0.88   \\
HE 0417-5357  &  2.19 &  8.48  & 2000.07.15 & 10:16:58 &  -8.104 & -6.21 & 3.25   \\
              &       &        & 2000.07.18 & 09:37:59 &  -8.014 &  6.47 & 2.31   \\
HE 0420-4748  & 57.29 &  0.06  & 2000.07.15 & 10:30:58 &  -6.686 & 57.37 & 2.83   \\
              &       &        & 2000.07.18 & 10:19:41 &  -6.498 & 57.26 & 1.52   \\
HE 0423-1434  & 55.50 &  5.98  & 2002.09.23 & 06:50:16 &   4.515 & 45.45 & 4.07   \\
              &       &        & 2002.12.07 & 08:17:12 & -22.990 & 57.28 & 1.71   \\
WD 0615-591   &  1.94 &  0.61  & 2000.09.14 & 09:26:25 & -13.891 & 1.322 & 2.83   \\%
              &       &        & 2001.04.08 & 01:04:56 & -21.666 & 2.251 & 2.00   \\%
WD 0845-188   & 72.10 &  5.00  & 2001.12.18 & 07:32:37 &  2.566  & 70.29 & 1.79   \\%
              &       &        & 2001.12.29 & 07:17:47 & -0.614  & 79.02 & 3.50   \\%
WD 1252-289   & 16.99 &  0.59  & 2000.04.21 & 04:49:17 & -8.262  & 16.83 & 1.46   \\
              &       &        & 2000.05.19 & 01:56:15 & -20.292 & 18.10 & 3.85   \\
WD 1326-037   & 51.82 &  6.61  & 2001.01.14 & 09:11:41 &  33.888 & 59.85 & 1.91   \\
              &       &        & 2001.02.02 & 09:27:35 &  31.681 & 49.09 & 1.11   \\
WD 1428-125   & 48.55 &  3.51  & 2000.07.06 & 02:51:29 & -20.736 & 53.20 & 5.01   \\
              &       &        & 2000.07.13 & 02:26:45 & -22.017 & 47.22 & 2.68   \\
WD 1445+152   & -4.35 &  3.48  & 2002.04.23 & 05:32:37 &  11.695 & -4.35 & 1.39   \\
WD 1542+182   & 27.89 &  4.49  & 2002.04.23 & 08:22:15 &  20.960 & 27.89 & 1.80   \\
WD 1612-111   & 45.50 &  2.48  & 2000.06.06 & 06:35:11 &   6.461 & 47.37 & 2.68   \\
              &       &        & 2000.06.08 & 02:30:50 &   5.989 & 43.85 & 2.52   \\
WD 2144-079   &  9.97 &  3.90  & 2000.05.17 & 09:50:49 &  40.410 & 11.93 & 1.75   \\
              &       &        & 2000.05.19 & 09:58:26 &  40.360 & 6.080 & 2.47   \\
WD 2234+064   & 19.21 &  4.19  & 2001.06.18 & 09:59:17 &  37.415 & 14.75 & 4.18   \\
              &       &        & 2001.09.01 & 01:22:15 &  12.083 & 17.27 & 2.54   \\
              &       &        & 2002.09.24 & 04:30:27 &   0.547 & 23.40 & 2.75   \\
WD 2354+159   & 44.02 &  1.21  & 2002.09.04 & 06:44:23 &  18.606 & 43.46 & 1.43   \\
              &       &        & 2002.09.13 & 04:25:56 &  14.692 & 45.33 & 2.19   \\
WD 2354-305   & 34.30 & 10.00  & 2000.07.15 & 05:28:58 &  20.072 & 34.30 & 4.01
\enddata
\end{deluxetable}

\begin{deluxetable}{lccccccc}
\tablewidth{0pt}
\tabletypesize{\footnotesize}
\tablecaption{Biweight Apparent Velocities\label{results_mix}}
\tablehead{
\colhead{Sample} & \colhead{Number of WDs} & \colhead{$\langle v_{\rm app}\rangle_{\rm BI}$} & \colhead{$\delta\langle v_{\rm app}\rangle_{\rm BI}$} & \colhead{$\sigma(v_{\rm app})_{\rm BI}$} & \colhead{$\langle \delta$$v_{\rm app}\rangle$} & \colhead{$\langle M/R\rangle_{\rm BI}$} & \colhead{$\delta\langle M/R\rangle_{\rm BI}$}
\\
\colhead{} & \colhead{} & \colhead{(km s$^{-1}$)} & \colhead{(km s$^{-1}$)} & \colhead{(km s$^{-1}$)} & \colhead{(km s$^{-1}$)} &\colhead{($M_\odot/R_\odot$)} & 
\colhead{($M_\odot/R_\odot$)}
}
\startdata
DBA                           & 16  & 40.8   & 4.7   & 16.6   & 7.1   & 64.2   &  7.4    \\
DBA (He)\tablenotemark{a}     & 16  & 34.0   & 5.0   & 18.6   & 4.7   & 53.4   &  7.8    \\
DB (He)\tablenotemark{b}      & 20  & 42.9   & 8.9   & 23.9   & 4.1   & 67.4   & 13.9    \\
DBA+DB (He)\tablenotemark{b}  & 36  & 43.4   & 7.9   & 22.4   & 4.5   & 68.9   & 12.5    \\
DA\tablenotemark{c}          & 449 & 32.57  & 1.17  & 24.84  & 1.51  & 51.19  & 1.84
\enddata
\tablenotetext{a}{Sample before pressure shift correction described in Section \ref{heline}.}
\tablenotetext{b}{Sample with applied pressure shift correction.}
\tablenotetext{c}{Main sample of normal DAs from \citet{Falcon10}.  The values listed in this row contain arithmetic means, standard deviation, and uncertainties and no biweights.}
\end{deluxetable}

\begin{deluxetable}{lccccccc}
\tablewidth{0pt}
\tabletypesize{\footnotesize}
\tablecaption{Biweight Masses\label{table_mass}}
\tablehead{
\colhead{Sample} & \colhead{Number of WDs} & \colhead{$\langle v_{\rm app}\rangle_{\rm BI}$} & \colhead{$\delta$\,$\langle v_{\rm app}\rangle_{\rm BI}$} & \colhead{$\langle T_{\rm eff}\rangle_{\rm BI}$} & \colhead{$\sigma(T_{\rm eff})_{\rm BI}$} & \colhead{$\langle M\rangle_{\rm BI}$} & \colhead{$\delta$\,$\langle M\rangle_{\rm BI}$}
\\
\colhead{} & \colhead{} & \colhead{(km s$^{-1}$)} & \colhead{(km s$^{-1}$)} & \colhead{(K)} & \colhead{(K)} & \colhead{($M_\odot$)} & 
\colhead{($M_\odot$)}
}
\startdata
DBA          & 16 & 40.8   & 4.7   & 17890 & 1040 & 0.71  & $^{+0.04}_{-0.05}$   \\
DB (He)      & 20 & 42.9   & 8.9   & 20570 & 3280 & 0.74  & $^{+0.08}_{-0.09}$   \\
DBA+DB (He)  & 36 & 43.4   & 7.9   & 19260 & 2770 & 0.74  & $^{+0.07}_{-0.08}$   \\
DA\tablenotemark{a} & 449 & 32.57 & 1.17 & 19400 & 9950 & 0.647 & $^{+0.013}_{-0.014}$
\enddata
\tablenotetext{a}{Main sample of normal DAs from \citet{Falcon10}.  The values listed in this row contain arithmetic means, standard deviation, and uncertainties and no biweights.}
\end{deluxetable}

\end{document}